\documentclass[twocolumn,aps,prd,10pt,showpacs]{revtex4-2}
\usepackage{bm}
\usepackage{amsfonts}
\usepackage{amssymb}
\usepackage{amsmath}
\usepackage{bm}
\usepackage{epsfig}
\begin{document}
\title{New solutions of the Dirac, Maxwell and Weyl equations\\ from the fractional Fourier transform}
\author{Iwo Bialynicki-Birula}\email{birula@cft.edu.pl}
\affiliation{Center for Theoretical Physics, Polish Academy of Sciences\\
Aleja Lotnik\'ow 32/46, 02-668 Warsaw, Poland}
\begin{abstract}
New solutions of relativistic wave equations are obtained in a unified manner from generating functions of spinorial variables. The choice of generating functions as Gaussians leads to representations in the form of generalized fractional Fourier transforms. Wave functions satisfying the Dirac, Maxwell, and Weyl equations are constructed by simple differentiations with respect to spinorial arguments. In the simplest case, one obtains Maxwell and Dirac hopfion solutions.
\end{abstract}

\maketitle

\section{Introduction}

The aim of this work is to describe a new universal tool in the study of the solutions of Weyl, Maxwell and Dirac equations. This tool is used here to derive a large family of solutions of these equations. The essential role is played by the generating function $\Upsilon(\eta,\eta^*|x)$ which in a simple form encodes the information about the solutions of Weyl and Maxwell equations and by the function $\Upsilon(\eta,\eta^*,\zeta,\zeta^*|x)$ which encodes the information about the solutions of the Dirac equation. Solutions of relativistic wave equations are obtained by the differentiation of generating functions with respect to their spinorial arguments $\eta$ and $\zeta$.

The generating functions have the form of generalized fractional Fourier transforms. Fractional Fourier transform (sometimes also called the Fresnel transform) was introduced long time ago by Condon \cite{con}. However, its numerous applications in optics, in signal processing, in image compression, in computed tomography, and other fields began after the work of Namias \cite{nam}. The generalization of the fractional Fourier transform to spinorial variables extends its range of applications. It enables one to obtain analytic localized solutions of relativistic wave equations with ever increasing degree of complexity. The term {\em localized} has here the following meaning. The probability to find the particle outside the sphere of a fixed radius tends to zero when the scaling parameter $a$ tends to zero. Of course, the localization can be achieved only around $t=0$ because {\em all} solutions of wave functions describing particles evolving in free space undergo dispersion. As a result, the mean square radius is a quadratic function of time \cite{rs,bb1},
\begin{align}\label{disp}
\langle{\bm r}^2\rangle_t=A+Bt^2.
\end{align}
Therefore, in the remote past and future the extension of {\em every} wave packet is arbitrarily large.

The generating functions introduced here depend on complex parameters: the components of the relativistic spinors and their complex conjugates. These generating functions have interesting properties on their own because they form a representations of the full 15 parameter conformal group for Weyl and Maxwell equations and of the inhomogeneous Poincar\'e group for the Dirac equation. Our method of generating solutions is particularly useful for the construction of knotted solutions with intricate topological properties. As special cases one obtains the solutions describing the Dirac \cite{bb1} and Maxwell \cite{syn,ran,kl} hopfions.

The main mathematical ingredients in this work are two-component spinors and four component bispinors appearing as arguments of the functions that generate solutions of Weyl, Maxwell and Dirac equations. These solutions are obtained as the derivatives of the generating functions with respect to their spinorial arguments. Of course, spinors and bispinors also appear as the wave functions in spacetime. Our notation is a slight modification of that used in \cite{cors,pr1,bb2} and it is summarized in the Appendix A.

The formalism based on spinors is particularly well suited to describe the solutions of relativistic wave equations because it is often directly connected with topological properties of these solutions. These properties in many cases \cite{ran,kl} involve Hopf fibration \cite{hopf}. This subject in the case of the electromagnetic field has been thoroughly studied (see a recent review \cite{abt}). The spinorial representation introduced in the present work gives a unified framework to describe also the solutions of other relativistic wave equations.

\section{Massless particles: Weyl and Maxwell equations}

Let us consider an arbitrary complex function $\Upsilon(\eta,\eta^*)$ of a two-component spinor $\eta^A$ and its complex conjugate $\eta^{\dot{A}}$. The space of these functions becomes a representation of the conformal group after the introduction of the 15 group generators. The generators of translations $P_\mu$, rotations and special Lorentz transformations $M_{\mu\nu}$, special conformal transformations $K_\mu$, and dilation $D$ are ($\hbar=1, c=1$):
\begin{subequations}
\begin{align}\label{poincm}
P_\mu&=-\partial_{\dot{A}}g_\mu^{\;\;\dot{A}B}\partial_B,\\
M_{\mu\nu}&=\frac{i}{2}\left(\eta^AS_{\mu\nu A}^{\quad\;\;B}\partial_B-
\eta^{\dot{A}}S_{\mu\nu\dot{A}}^{\quad\;\;\dot{B}}\partial_{\dot{B}}\right),\\
K_\mu&=\eta^Ag_{\mu A\dot{B}}\eta^{\dot{B}},\\
D&=\frac{1}{2i}\left(\eta^A\partial_A+\eta^{\dot{A}}\partial_{\dot{A}}\right),
\end{align}
\end{subequations}
where $\partial_{A}=\partial/\partial\eta^A$ and $\partial_{\dot{A}}=\partial/\partial\eta^{\dot{A}}$.
This representation has a close connection with twistors \cite{pm,ws,bb3} but this line of investigation will not be pursued here.

The generators of translations are of special significance because with their help one can construct from the function $\Upsilon(\eta,\eta^*)$ a complete field defined at all spacetime points $x^\mu$. To this end, $\Upsilon(\eta,\eta^*)$ is represented as a four-dimensional Fourier integral over the spinorial variables,
\begin{align}\label{sint}
\Upsilon(\eta,\eta^*)\!=\!\!
\int\!\!d^2\!\kappa d^2\!\kappa^*\!\exp\left[-i(\kappa_A\eta^A+\eta^{\dot{A}}\kappa_{\dot{A}})\right]
{\tilde\Upsilon}(\kappa,\kappa^*).
\end{align}
The integration variables are the real and imaginary parts of both components of the spinor $\kappa$. The translation operators acting on the spinorial integral (\ref{sint}) produce an integral in the form of a multidimensional fractional Fourier transform since the exponent has both the quadratic and the linear part in the integration variables,
\begin{widetext}
\begin{align}\label{sx}
\Upsilon(\eta,\eta^*|x)
=\exp(-iP_\mu x^\mu)\Upsilon(\eta,\eta^*)
=\int\!d^2\!\kappa d^2\!\kappa^*
\exp\left[- i\kappa_{\dot{A}}
g_\mu^{\;\;\dot{A}B}\kappa_Bx^\mu\right]
\exp\left[-i(\kappa_A\eta^A+\eta^{\dot{A}}\kappa_{\dot{A}})\right]
{\tilde\Upsilon}(\kappa,\kappa^*).
\end{align}
\end{widetext}
This function satisfies four Schr\"odinger-like equations,
\begin{align}\label{sch}
i\partial_\mu\Upsilon(\eta,\eta^*|x)=P_\mu\Upsilon(\eta,\eta^*|x).
\end{align}
For every choice of ${\tilde\Upsilon}(\kappa,\kappa^*)$ which guarantees the convergence of the integral and for all values of the spinorial parameters $\eta$ and $\eta^*$, the generating function $\Upsilon(\eta,\eta^*|x)$ satisfies the d'Alembert equation,
\begin{align}\label{dal}
(\partial_t^2-\Delta)\Upsilon(\eta,\eta^*|x)=0,
\end{align}
because the derivative $\partial_\mu$ produces $\kappa_{\dot{A}}g_\mu^{\;\;\dot{A}B}\kappa_B$ under the integral and this is a lightlike vector. Negative energy solutions are obtained by the translation $\exp(iP_\mu x^\mu)$. The general solution is a superposition of positive and negative energy contributions but choosing only one sign at a time simplifies the formulas.

The generation of the solutions of Maxwell equations from a solution of the d'Alembert equation has been known already to Whittaker \cite{whitt}. This method was formulated in the spinorial framework by Penrose \cite{pen}. In both these constructions, the Maxwell field is built from second derivatives with respect to {\em spacetime variables}. Therefore, these methods are closely related to Hertz potentials. In contrast, the generation of the solutions of Weyl and Maxwell equations from our spinorial generating function $\Upsilon(\eta,\eta^*|x)$ is quite different because it involves the derivatives with respect to the auxiliary spinorial argument $\eta^A$. Derivatives with respect to the components of spinors also appear but in an entirely different role in the solutions of the massless wave equations expressed in terms of Penrose transforms \cite{pr2}.

The first derivative is a solution of the Weyl equation,
\begin{align}\label{weyl}
\phi_{C}(x)=i\partial_C\Upsilon(\eta,\eta^*|x),\quad
g^{\mu\dot{E}C}\partial_\mu\phi_{C}(x)=0.
\end{align}
The second derivative is a solution of the Maxwell equations,
\begin{align}\label{max} \phi_{CD}(x)=-\partial_C\partial_D\Upsilon(\eta,\eta^*|x),\quad
g^{\mu\dot{E}C}\partial_\mu\phi_{CD}(x)=0.
\end{align}
Both equations follow from the algebraic relation:
\begin{align}\label{arel}
g^{\mu\dot{E}C}\kappa_{\dot{A}}g_\mu^{\;\;\dot{A}B}\kappa_B\kappa_C
=2\kappa_{\dot{A}}\epsilon^{\dot{A}\dot{E}}\epsilon^{BC}\kappa_B\kappa_C=0,
\end{align}
applied to the derivatives of the spinorial transform (\ref{sx}). The wave equation (\ref{max}) is equivalent to Maxwell equations upon the following identification of the components of $\phi_{CD}(x)$ with the components of the Riemann-Silberstein (RS) vector \cite{rs}:
\begin{align}\label{rs}
F_x=\phi_{11}-\phi_{00},\;F_y=-i(\phi_{11}+\phi_{00}),
\;F_z=2\phi_{01}=2\phi_{10}.
\end{align}
It follows from these relations that positive/negative frequency solutions of Maxwell equations ${\bm F}^\pm(x)$ have the following spinorial representation:
\begin{align}\label{rs1}
{\bm F}^\pm(\bm r,t)&=\int\!d^2\!\kappa d^2\!\kappa^*
\left[\begin{array}{c}\kappa_1^2-\kappa_0^2\\
-i\kappa_1^2-i\kappa_0^2\\
2\kappa_0\kappa_1\end{array}\right]\nonumber\\
&\times\exp\left[\mp i\kappa_{\dot{A}}
g_\mu^{\;\;\dot{A}B}\kappa_Bx^\mu\right]
{\tilde\Upsilon}^\pm(\kappa,\kappa^*).
\end{align}
General solutions are sums of positive and negative frequency solutions. It is shown in the next Section that the spinorial representation through the Hopf fibration is directly related to the Fourier representation that is commonly used in physical applications. The alternative representation based on the Penrose transform \cite{pr2} does not have this property.

\section{Spinorial representation and\\the Hopf fibration}

The Hopf fibration \cite{hopf} is a decomposition of the three-dimensional sphere into linked circles: the fibers. Every fiber corresponds to one point on the two-dimensional sphere. In the Hopf construction the spheres are parametrized in terms of the Cartesian coordinates subjected to the condition that the radius of the sphere is fixed. The mapping of the points in the three-dimensional sphere $\xi_i$ onto the points in the two-dimensional sphere $k_i$ was defined by Hopf as follows:
\begin{align}\label{hopf}
k_x=2(\xi_1\xi_3&+\xi_2\xi_4),
\;k_y=2(\xi_2\xi_3-\xi_1\xi_4),\nonumber\\
k_z&=\xi_1^2+\xi_2^2-\xi_3^2-\xi_4^2.
\end{align}
In order to introduce the physical interpretation, the symbols in these formulas are different than those used by Hopf. The relations invented by Hopf preserve the length. The 3D sphere of unit radius is mapped onto the 2D sphere of unit radius since  $k_x^2+k_y^2+k_z^2=(\xi_1^2+\xi_2^2+\xi_3^2+\xi_4^2)^2$. The connection of the Hopf fibration with spinors is revealed when the parameters $\xi_i$ are identified with the real and imaginary parts of the spinor components $\kappa_0=\xi_1+i\xi_2,\,\kappa_1=\xi_3+i\xi_4$.

The physical content of the Hopf formula (\ref{hopf}) is best described with the help of Pauli matrices and it has the form of the relation between spinors $\kappa$ and the lightlike wave vectors: $k^\mu=\kappa_{\dot{A}}g^{\mu{\dot{A}}B}\kappa_B$. The space components of the wave vector are the Hopf parameters  $k_x,k_y,k_z$. The fibers are formed by those spinors that differ only by an overall phase factor $e^{i\varphi}$. This phase factor is not uniquely defined. One may choose, for example, the phase of the upper spinor component. In this case  $\varphi=\arctan(\xi_2/\xi_1)$. With this choice, the relations (\ref{hopf}) can be inverted,
\begin{align}\label{hopf1}
\xi_1=\frac{\sqrt{k_z+k}\cos\varphi}{\sqrt{2}},\;
\xi_2=\frac{\sqrt{k_z+k}\sin\varphi}{\sqrt{2}},\nonumber\\
\xi_3=\frac{k_x\cos\varphi-k_y\sin\varphi}{\sqrt{2}\sqrt{k+k_z}},\;
\xi_4=\frac{k_y\cos\varphi+k_x\sin\varphi}{\sqrt{2}\sqrt{k+k_z}},
\end{align}
where $k=\sqrt{k_x^2+k_y^2+k_z^2}$.
The relation between the spinorial representation (\ref{rs1}) and the Fourier representation of the electromagnetic field represented by the RS vector \cite{rs,qm,qnum},
\begin{align}\label{frep}
{\bm F}(\bm r,t)&=\int\!\!\frac{d^3k}{(2\pi)^{3/2}}\left[\!\begin{array}{c}
-k_xk_z+ikk_y\\
-k_yk_z-ikk_x\\
k_x^2+k_y^2\end{array}\right]\nonumber\\
&\times\left[f_+(\bm k)e^{i\bm k\cdot\bm r-i\omega t}+f_-^*(\bm k)e^{-i\bm k\cdot\bm r+i\omega t}\right],
\end{align}
is obtained by the change of variables. The integration with respect to two components of the spinor in (\ref{rs1}) is equivalent to the integration over the three components of the wave vector and an additional integration with respect to the phase $\varphi$. The Jacobian of the transformation $\xi_i\to\{k_x,k_y,k_z,\varphi\}$ is equal to $1/8k$ and the integral (\ref{rs1}) expressed in terms of new variables coincides with (\ref{frep}) provided we make the following identification:
\begin{subequations}
\begin{align}\label{rs3}
f_+(\bm k)=\left(\frac{\pi}{2}\right)^{3/2}
\!\!\!\int_0^{2\pi}\!\!\!\!d\varphi\,\frac{e^{2i\varphi}{\tilde\Upsilon}^+({\bm k},\varphi)}{k(k_x-ik_y)},\\
f_-^*(\bm k)=\left(\frac{\pi}{2}\right)^{3/2}
\!\!\!\int_0^{2\pi}\!\!\!\!d\varphi\,\frac{e^{2i\varphi}{\tilde\Upsilon}^-({\bm k},\varphi)}{k(k_x-ik_y)}.
\end{align}
\end{subequations}
Even though the spinorial representation and the Fourier representation are equivalent, the spinorial representation is much easier to use in the derivation of various hopfion-like solutions.

\section{The hopfion family of solutions of Weyl and Maxwell equations}

A simple choice of ${\tilde\Upsilon}(\kappa,\kappa^*)$ that leads to the analytic solution is a Gaussian,
\begin{align}\label{gaussm}
{\tilde\Upsilon}(\kappa,\kappa^*)
=\exp\left(-\kappa_{\dot{A}}g_\mu^{\dot{A}B}\kappa_B\,a^\mu\right),
\end{align}
where $a^\mu$ is any complex vector. Without loss of generality, one may assume that $a^\mu$ is a real vector because the imaginary part of $a^\mu$ can be eliminated by a shift of the origin of coordinate system. From now on it is assumed that the coordinate system is chosen in such a way that $a^\mu=\{a,0,0,0\}$ although in some formulas the full vector $a^\mu$ will appear. In order to guarantee the convergence of the integrals, it is assumed that $a>0$. In this case the integration can be easily done and after dropping the irrelevant factor $\pi^2$, one obtains,
\begin{align}\label{res}
\Upsilon(\eta,\eta^*|x)&=D(x)\exp\left(-iD(x)\eta^Ag_{\mu A\dot{B}}(x^\mu-ia^\mu)\eta^{\dot{B}}\right),
\end{align}
where $D(x)=((a+it)^2+x^2+y^2+z^2)^{-1}$. This simple calculation shows the advantage of using the spinorial representation. The evaluation of the corresponding integrals in the Fourier representation would have been much more complicated. This method of generating solutions is applicable also to wave equations for higher spins. In particular, the gravitational waves in linearized gravity are described by fourth derivatives of the generating function.

The solutions of Weyl and Maxwell equations obtained from the function (\ref{res}) by differentiation, according to the formulas (\ref{weyl}) and (\ref{max}), have the form:
\begin{align}
\phi_{C}(x)&=D(x)\exp\left(-\eta^A\psi_A(x)\right)
\psi_C(x),\label{res1}\\
\phi_{CD}(x)&=D(x)\exp\left(-\eta^A\psi_A(x)\right)
\psi_C(x)\psi_D(x),\label{res2}
\end{align}
where
\begin{align}\label{psi}
\psi_A(x)=D(x)g_{\mu A\dot{B}}\eta^{\dot{B}}(x^\mu-ia^\mu).
\end{align}
These formulas contain arbitrary parameters $\eta^A$ and $\eta^{\dot{A}}$. By differentiating (\ref{res1}) and  (\ref{res2}) (or integrating with any function of these parameters) with respect to $\eta^A$ and/or $\eta^{\dot{A}}$ one still obtains solutions of the wave equations.

One may check by a direct calculation that not only for the exponential functions appearing in (\ref{res1}) and (\ref{res2}) but for all functions $h\left(\psi(x)\right)$ the spinors
\begin{align}
\phi_{C}(x)&=D(x)h_1\left(\psi(x)\right)
\psi_C(x),\label{res2w}\\
\phi_{CD}(x)&=D(x)h_2\left(\psi(x)\right)
\psi_C(x)\psi_D(x)\label{res2m}
\end{align}
are solutions of the wave equations (\ref{weyl}) and (\ref{max}). One obtains in this way a large class of solutions of Weyl and Maxwell equations controlled by an arbitrary complex function of the two components of $\psi_A(x)$.

Incidentally, one obtains a realization of de Broglie idea of fusion \cite{db} by choosing $h_2\left(\psi(x)\right)=h_1\left(\psi(x)\right) h_1\left(\psi(x)\right)$. Indeed, it looks like ``photons are made of two neutrinos'' because the photon wave function (apart from the factor $D(x)$) is a product of neutrino wave functions, $\phi_{AB}=D(x)^{-1}\phi_{A}\phi_{B}$.

The simplest hopfion solutions of the Weyl equation are obtained from (\ref{res2w}) when $h(\psi)=-i$ and by choosing either $\eta^{\dot{A}}=\{1,0\}$ or $\eta^{\dot{A}}=\{0,1\}$.
\begin{align}\label{hopfw}
\phi^1_{A}(x)=\left[D(x)\right]^2
\left[\begin{array}{c}t_+\\
 x_+\end{array}\right],\; \phi^2_{A}(x)=\left[D(x)\right]^2
\left[\begin{array}{c}x_-\\
 t_- \end{array}\right],
\end{align}
where $t_\pm=t\pm z-ia$ and $x_\pm=x\pm iy$.

Owing to the appearance of the product of spinors in (\ref{res2m}), the electromagnetic field given by this formula is null; both field invariants vanish, i.e., ${\bm E}^2-{\bm B}^2=0$ and ${\bm E}\cdot{\bm B}=0$. Null fields play a special role in electromagnetism. They possess intriguing topological properties \cite{kl}. The simplest solutions are obtained from the formula (\ref{res2m}) by choosing the same spinors $\psi_A$ as for the Weyl equation. The two closely related RS vectors constructed from these spinors according to the formulas (\ref{rs}) and (\ref{res2m}) are:
\begin{subequations}
\begin{align}\label{hopf2}
{\bm F}^1_H=\left[D(x)\right]^3\left[\begin{array}{c}
t_+^2-x_+^2\\
i(t_+^2+x_+^2)\\
-2t_+x_+
\end{array}\right],\\
{\bm F}^2_H=\left[D(x)\right]^3\left[\begin{array}{c}
-t_-^2+x_-^2\\
i(t_-^2+x_-^2)\\
-2x_-t_-
\end{array}\right].
\end{align}
\end{subequations}
The first formula coincides with Eq.~(23) of \cite{qnum}. The electric and magnetic field vectors (i.e., the real and imaginary parts of ${\bm F}_H$) describe the simplest knotted solutions of Maxwell equations: the hopfion. It was discovered by Synge \cite{syn} who interpreted it as ``an electromagnetic model of a material particle''. Its intricate topological properties were discovered by Ra$\tilde{\rm n}$ada \cite{ran} who found the connection with Hopf fibration.
\begin{figure}[b]
\centering
\includegraphics[width=4cm,height=4cm]{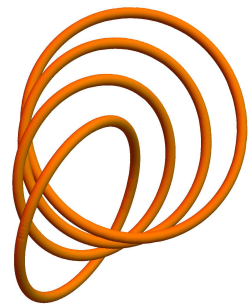}
\caption{Linked circles: the trademark of the Hopf fibration. The lines of velocity plotted here are obtained by solving the set of differential equations $d{\bm r}(\lambda)/d\lambda={\bm v}({\bm r}(\lambda))$ for different initial conditions.}
\label{fig1}
\end{figure}
The hopfion solution can be obtained in many different ways, even from a simple Fourier integral \cite{qnum}. However, a method of choice to obtain also other solutions with even more intricate topological properties is the Bateman construction \cite{kl,bat}. Bateman discovered that if two complex functions of spacetime variables $\alpha(x,y,z,t)$ and $\beta(x,y,z,t)$ obey the condition;
\begin{align}\label{bat}
\nabla\alpha\times\nabla\beta
=i(\partial_t\alpha\nabla\beta-\partial_t\beta\nabla\alpha),
\end{align}
then the vector $F_B=\nabla\alpha\times\nabla\beta$ is a (null) solution of Maxwell equations.

The Bateman construction is mentioned here because there is a direct connection between the spinorial method and this construction. Namely, the two components of the spinor (\ref{psi}) can be used as $\alpha$ and $\beta$ in the Bateman construction because they obey the condition (\ref{bat}). The RS vector obtained from the Bateman construction differs only by the factor $-i$ from the one obtained from (\ref{rs}) and (\ref{hopf2}). All solutions with intricate topological properties, analyzed in \cite{kl}, can be obtained from the formula (\ref{res2m}) by choosing the function $h(\psi_A)$ in the form $h(\psi_A)=(\psi_0(x))^p(\psi_1(x))^q$, where $p$ and $q$ are relatively prime integers.

The solutions of Weyl and Maxwell equations (\ref{res2w}) and (\ref{res2m}) have one common feature. In both cases we find for all choices of the function $h$ the same lightlike four-vector $l^\mu=\psi_{\dot{A}}g^{\mu{\dot{A}B}}\psi_B$ characterizing the solution. The current $j^\mu$ for the solutions of the Weyl equation and the energy-momentum tensor $T^{\mu\nu}$ for the solutions of the Maxwell equations are built from $l^\mu$,
\begin{subequations}
\begin{align}\label{cen}
j^\mu &=|D(x)h_1\left(\psi(x)\right)|^2l^\mu(x),\\
T^{\mu\nu}&=|D(x)h_2\left(\psi(x)\right)|^2l^\mu(x)l^\nu(x).
\end{align}
\end{subequations}
The properties of the vector $l^\mu(x)$ underscore the connection with Hopf fibration. Namely, integral lines of velocity $\bm v=\bm l/l^0$ form linked circles, as shown in Fig.~1, which is a characteristic feature of Hopf fibration.

\section{Massive particles: Dirac equation}

The standard Dirac equation,
\begin{align}\label{dir}
\left(i\gamma^\mu\partial_\mu-m\right)\Psi(x)=0,
\end{align}
is converted in the Weyl representation \cite{weyl} of $\gamma$ matrices,
\begin{align}\label{weylm}
\gamma^\mu=\left[\begin{array}{cc}0&g^\mu_{\;\;A\dot{B}}\\g^{\mu\dot{A}B}&0
\end{array}\right],\quad\Psi(x)=\left[\begin{array}{c}\phi_A(x)\\
\chi^{\dot{A}}(x)\end{array}\right],
\end{align}
into the following set of two equations for two-component spinors:
\begin{align}\label{dirweyl}
ig^{\mu\dot{A}B}\partial_\mu\phi_B(x)=m\chi^{\dot{A}}(x),\quad
ig^\mu_{\;\;A\dot{B}}\partial_\mu\chi^{\dot{B}}(x)=m\phi_{A}(x).
\end{align}
Due to the presence of two spinors in these equations, the generating function $\Upsilon$ should have two spinorial arguments. The appearance of mass calls for some modifications of the spinorial formalism. The following spinorial integral, patterned after the integral (\ref{sx}) for the solutions of d'Alembert equation, produces positive energy solutions of the Klein-Gordon equation,
\begin{widetext}
\begin{align}\label{kg}
\Upsilon(\eta,\eta^*,\zeta,\zeta^*|x)
&=\int\!d^2\!\kappa d^2\!\kappa^* d^2\!\lambda d^2\!\lambda^*
\exp\left[-i(\kappa_A\eta^A+\eta^{\dot{A}}\kappa_{\dot{A}}
+\lambda^A\zeta_A+\zeta_{\dot{A}}\lambda^{\dot{A}})\right]
\exp\left[-i\left(\kappa_{\dot{A}}g_\mu^{\;\;\dot{A}B}\kappa_B
+\lambda^Ag_{\mu\dot{B}}\lambda^{\dot{B}}\right)x^\mu\right]\nonumber\\
&\times\delta\left(\kappa_A\lambda^A+\kappa_{\dot{A}}
\lambda^{\dot{A}}-m\right)
\delta\left(i(\kappa_{\dot{A}}\lambda^{\dot{A}}-\kappa_A\lambda^A)\right)
{\tilde\Upsilon}(\kappa,\kappa^*,\lambda,\lambda^*),
\end{align}
\end{widetext}
because the action of the d'Alembertian produces under the integral sign the following expression:
\begin{align} \kappa_{\dot{A}}g^{\mu\dot{A}C}\kappa_C\lambda^Ag_{\mu A\dot{B}}\lambda^{\dot{B}}
=4\kappa_A\lambda^A\kappa_{\dot{A}}\lambda^{\dot{A}}.
\end{align}
The presence of the $\delta$-functions enables one to replace this expression by $m^2$, resulting in the Klein-Gordon equation,
\begin{align}\label{kg1}
\left(\partial^2_t-\Delta+m^2\right)
\Upsilon(\eta,\eta^*,\zeta,\zeta^*|x)=0.
\end{align}
Negative energy solutions are obtained by reversing the sign of $x^\mu$. In full analogy with the massless case, the solutions of the Dirac equation are obtained from (\ref{kg}) by differentiation with respect to spinorial parameters $\eta^C$ and $\zeta_{\dot{C}}$. The Dirac bispinor $\Psi(x)$ has the form:
\begin{align}\label{dhopf}
\Psi(x)=\left[\begin{array}{c}
i\partial_A\Upsilon(\eta,\eta^*,\zeta,\zeta^*|x)\\
i\eth^{\dot{A}}\Upsilon(\eta,\eta^*,\zeta,\zeta^*|x)
\end{array}\right],
\end{align}
where $\partial_A=\partial/\partial\eta^A$ and $\eth^{\dot{A}}=\partial/\partial\zeta_{\dot{A}}$.

In order to obtain explicit formulas for the solutions, we must choose the function ${\tilde\Upsilon}(\kappa,\kappa^*,\lambda,\lambda^*)$ in such a way that the integrations can be performed.

\section{The hopfion family of solutions of the Dirac equation}

The existence of similar representations of the solutions of Maxwell and Dirac equations enables one to define a map between these solutions. Namely, by choosing ${\tilde\Upsilon}(\kappa,\kappa^*,\lambda,\lambda^*)$  as a product of functions appearing in (\ref{sx}),
\begin{align}\label{prod}
{\tilde\Upsilon}(\kappa,\kappa^*,\lambda,\lambda^*)
={\tilde\Upsilon}_1(\kappa,\kappa^*)
{\tilde\Upsilon}_2(\lambda,\lambda^*),
\end{align}
one establishes a direct relation between solutions of Maxwell and Dirac equations. To every pair of solutions of Maxwell equations there corresponds a solution of the Dirac equation. In particular, it is tempting to choose ${\tilde\Upsilon}(\kappa,\kappa^*)$ in the Gaussian form because this choice corresponds to the Maxwell hopfion.
\begin{subequations}
\begin{align}\label{gauss}
{\tilde\Upsilon}_1(\kappa,\kappa^*)
=&\exp\left[-\kappa_{\dot{A}}g_\mu^{\;\;\dot{A}B}\kappa_B\,a^\mu\right],\\
{\tilde\Upsilon}_2(\lambda,\lambda^*)
=&\exp\left[-\lambda^Ag_{\mu A\dot{B}}\lambda^{\dot{B}}\,a^\mu\right].
\end{align}
\end{subequations}

In order to do the calculations, the integral representations of the $\delta$-functions are introduced into the formula (\ref{kg}),
\begin{widetext}
\begin{align}\label{dx2}
&\Upsilon(\eta,\eta^*,\zeta,\zeta^*|x)
=\int\!dudv\!\int\!d^2\!\kappa d^2\!\kappa^* d^2\!\lambda d^2\!\lambda^*
\exp\left[-i(\kappa_A\eta^A+\eta^{\dot{A}}\kappa_{\dot{A}}
+\lambda^A\zeta_A+\zeta_{\dot{A}}\lambda^{\dot{A}})\right]\nonumber\\
&\times\exp\left[-\left(\kappa_{\dot{A}}g_\mu^{\;\;\dot{A}B}\kappa_B
+\lambda^Ag_{\mu A\dot{B}}\lambda^{\dot{B}}\right)(a^\mu+ix^\mu)\right]
\exp\left[-iu(\kappa_A\lambda^A
+\lambda^{\dot{A}}\kappa_{\dot{A}}-m)\right]
\exp\left[-v(\kappa_A\lambda^A
-\lambda^{\dot{A}}\kappa_{\dot{A}})\right],
\end{align}
\end{widetext}
where ${\tilde\Upsilon}(\kappa,\kappa^*,\lambda,\lambda^*)$ was replaced by the product of functions (\ref{prod}) and the irrelevant factor $(2\pi)^2$ was omitted. The integral with respect to the spinorial variables again has the form of the fractional Fourier transform. This integral can be evaluated and we are left with an integral with respect to $u$ and $v$,
\onecolumngrid
\begin{widetext}
\begin{figure}[t]
\centering
\includegraphics[width=17cm,height=4.7cm]{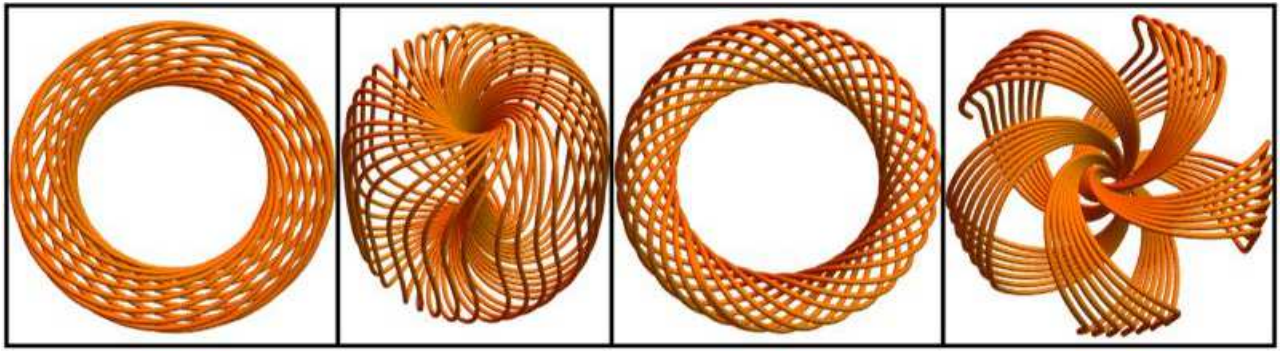}
\caption{The knotted lines of the current $\bm j={\bar\Psi}{\bm\gamma}\Psi$ for the solutions of the Dirac equation $\Psi_2,\Psi_4,\Psi_6$ and $\Psi_8$ listed in the Appendix B. The initial conditions are the same in all cases: $x_0=1,\,y_0=0.3,\,z_0=0$ and $a=1$. The distances are measured in electron Compton wave length and the size of the box is 3.}
\label{fig2}
\end{figure}
\begin{figure}[t]
\centering
\includegraphics[width=17cm,height=4.7cm]{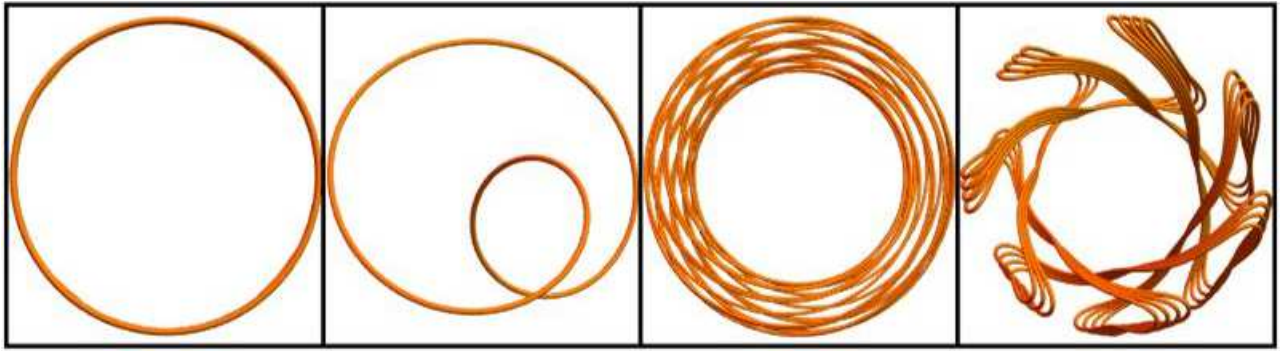}
\caption{Strong dependence on the initial conditions. The lines of current are plotted for the same solution $\Psi_4$ as in Fig.~2 but for different initial conditions for one of the coordinates, $x_0=\{0.25,0.6,0.1,1.25\}$.}
\label{fig3}
\end{figure}
\end{widetext}
\twocolumngrid
\begin{widetext}
\begin{align}\label{dx3}
\Upsilon(\eta,\eta^*,\zeta,\zeta^*|x)=&\int\!du\,dv\, e^{imu}[D(u,v|x)]^2\exp\left[-D(u,v|x)\left((\eta^Ag_{\mu A\dot{B}}\eta^{\dot{B}}+\zeta_{\dot{A}}g_\mu^{\;\;\dot{A}B}\zeta_B)
(a^\mu+ix^\mu)\right)\right]\nonumber\\
&\times\exp\left[-D(u,v|x)\left(-iu(\eta^A\zeta_A+\zeta_{\dot{A}}\eta^{\dot{A}})
-v(\eta^A\zeta_A-\zeta_{\dot{A}}\eta^{\dot{A}})\right)\right],
\end{align}
\end{widetext}
where $D(u,v|x)=\left[u^2+v^2+(a_\mu+ix_\mu)(a^\mu+ix^\mu)\right]^{-1}$.

This integral cannot be evaluated for arbitrary values of the spinorial parameters $\eta$ and $\zeta$. However, the expansion in powers of these parameters leads to integrals that can be explicitly evaluated. The simplest examples are the Dirac hopfions obtained by a different method in \cite{bb1}. The corresponding bispinors $\Psi_{\rm H}$ are obtained from the formula (\ref{dx3}) by evaluating second derivatives of $\Upsilon$ at the origin. The following four solutions of the Dirac equation are:
\begin{widetext}
\begin{subequations}\label{4sol}
\begin{align}
\eth^B\left[\begin{array}{c}
\partial_A\Upsilon\\
\eth^{\dot{A}}\Upsilon
\end{array}\right]_0
&=\int\!\!du\,dv\, e^{imu}[D(u,v|x)]^3\left[\begin{array}{c}
-(v+iu)\delta_A^B\\
g_\mu^{\;\;\dot{A}B}(a^\mu+ix^\mu)
\end{array}\right]=\frac{\pi m^2}{4}\left[\begin{array}{c}
\delta_A^B {\mathfrak K}_1\\
g_\mu^{\;\;\dot{A}B}(a^\mu+ix^\mu){\mathfrak K}_2
\end{array}\right],\\
\partial_{\dot{B}}\left[\begin{array}{c}
\partial_A\Upsilon\\
\eth^{\dot{A}}\Upsilon
\end{array}\right]_0
&=\int\!\!du\,dv\, e^{imu}[D(u,v|x)]^3\left[\begin{array}{c}
g_{\mu A\dot{B}}(a^\mu+ix^\mu)\\
(v-iu)\delta_{\dot{B}}^{\dot{A}}
\end{array}\right]=\frac{\pi m^2}{4}\left[\begin{array}{c}
g_{\mu A\dot{B}}(a^\mu+ix^\mu){\mathfrak K}_2\\
\delta_{\dot{B}}^{\dot{A}}{\mathfrak K}_1
\end{array}\right],
\end{align}
\end{subequations}
\end{widetext}
where ${\mathfrak K}_n$ are expressed in terms of the Macdonald functions ${\mathfrak K}_n=K_n(ms)/s^n$, and $s=\sqrt{(a+it)^2+x^2+y^2+z^2}$. The arguments of $\Upsilon$ and ${\mathfrak K}_n$ are omitted.

Choosing both values of the index ${\dot{B}}$ and both values of the index $B$ we obtain four Dirac hopfions corresponding to the formulas (7) and (8) of \cite{bb1} taken for the lowest values of the index $l$.
Higher order derivatives evaluated at the origin all give analytic solutions of the Dirac equation expressed in terms of Macdonald functions of increasing order. The integral lines of the current shown in Fig.~2 and Fig.~3 represent the solutions of the following three coupled differential equations:
\begin{align}\label{cur}
\frac{d{\bm r}(\lambda)}{d\lambda}={\bm j}({\bm r}(\lambda),t).
\end{align}
These figures were generated from the formulas in the Appendix B in the simplest case, when $t=0$. The lines of current depend strongly on the initial conditions, as shown in Fig.~3. All analytical calculations and plots were done with Mathematica \cite{wolf}.

\section{Conclusions}

Maxwell, Weyl, and Dirac equations play a fundamental role in relativistic quantum mechanics. The practically unlimited collection of new analytic solutions of these equations described here may help to understand better the intricate quantum properties of relativistic particles. The spinorial representation described here is particularly well suited in the analysis of solutions with intricate topological properties connected with Hopf fibration. The representation of the wave functions as derivatives with respect to spinorial variables makes their transformation properties transparent. Of course, it would also be possible to generate new solutions  by evaluating consecutive derivatives of some simple solution with respect to spacetime variables. However, this leads to highly complicated expressions. Already the second derivative of the simplest solution (\ref{d1}) produces a formula that is difficult to analyze because it is much more complicated than the expression (\ref{d7}) obtained by evaluating the seventh derivative with respect to spinorial variables.

\section{Acknowledgments}

I am thankful to Zofia Bialynicka-Birula for her very helpful criticism.

\appendix

\section{}
The two components of the spinor are labeled \cite{pr1} with 0 and 1. Spinors with upper and lower index are connected by the spinorial metric tensor $\epsilon$,
\begin{align}\label{not}
\phi^A=\epsilon^{AB}\phi_B,\;\phi_A=\psi^B\epsilon_{BA},\;
\epsilon_{AB}=\epsilon^{AB}
=\left[\begin{array}{cc}0\quad&1\\-1\quad& 0
\end{array}\right].
\end{align}
Repeated indices imply summation over two values of the index.
Under rotations and Lorentz transformations spinors are transformed by the unimodular matrices $S_A^{\;\;B}$
\begin{align}\label{tr}
'\phi_A=S_A^{\;\;B}\phi_B,\;\;'\phi^A=-S^A_{\;\;B}\phi^B.
\end{align}
The minus sign in the second formula is a consequence of the antisymmetry of the metric tensor and it implies the invariance of the scalar product $'\phi_A\,'\psi^A=\phi_A\psi^A$.

There are two inequivalent two-dimensional representations of the Lorentz group: the spinors $\phi_A$ and complex-conjugate spinors $\phi_{\dot{A}}$. Dotted indices signify complex conjugation, $\phi_{\dot{A}}=(\phi_A)^*$. The dotted spinors are transformed with the use of complex conjugate matrices,
\begin{align}\label{trc}
'\phi_{\dot{A}}=S_{\dot{A}}^{\;\;{\dot{B}}}\phi_{\dot{B}},\;\;
S_{\dot{A}}^{\;\;{\dot{B}}}=\left(S_A^{\;\;B}\right)^*.
\end{align}
Spinors with several indices transform as products of spinors for example,
\begin{align}\label{trc1}
'\phi_{\dot{A}B}=S_{\dot{A}}^{\;\;{\dot{C}}}S_B^{\;D}\phi_{\dot{B}D}.
\end{align}

An important role is played by spin-tensors---objects with tensorial and spinorial indices. There are four of them: $g^{\mu\dot{A}B},\;g^\mu_{\;\;A\dot{B}},\;S^{\mu\nu\;B}_{\;\;\,A}$ and $S^{\mu\nu\;{\dot{B}}}_{\;\;\,\dot{A}}$. They may all be expressed in terms of the Pauli matrices $\sigma_i$ and the $2\times 2$ unit matrix $I$,
\begin{subequations}
\begin{align}\label{pauli}
g^{0\dot{A}B}=I=g^0_{\;\;A\dot{B}},\;\;g^{i\dot{A}B}
=\sigma_i=-g^i_{\;\;A\dot{B}}\,,\\
S^{\mu\nu\;B}_{\;\;\,A}=
\frac{1}{2}\left(g^\mu_{\;\;A\dot{C}}g^{\nu\dot{C}B}
-g^\nu_{\;\;A\dot{C}}g^{\mu\dot{C}B}\right).
\end{align}
\end{subequations}
Spin-tensors are invariant under the simultaneous Lorentz transformations of vector and spinor indices.

\section{}

Selected solutions of the Dirac equation are obtained by evaluating the following derivatives of the generating function:
\begin{widetext}
\begin{align}\label{der}
\Psi_2=\eth^0\left[\begin{array}{c}
\partial_A\Upsilon\\
\eth^{\dot{A}}\Upsilon
\end{array}\right]_0,\:
&\Psi_4=\eth^0\eth^{\dot{0}}\eth^1\left[\begin{array}{c}
\partial_A\Upsilon\\
\eth^{\dot{A}}\Upsilon
\end{array}\right]_0,\;
\Psi_6=\partial^0\partial^{\dot{0}}\partial^1\partial^{\dot{1}}
\eth^{\dot{1}}\left[\begin{array}{c}
\partial_A\Upsilon\\
\eth^{\dot{A}}\Upsilon
\end{array}\right]_0,
&\Psi_8=\partial^{\dot{0}}\partial^1\partial^{\dot{1}}
\eth^0\eth^0\eth^{\dot{0}}\eth^{\dot{1}}\left[\begin{array}{c}
\partial_A\Upsilon\\
\eth^{\dot{A}}\Upsilon
\end{array}\right]_0.
\end{align}
All derivatives are to be taken at the origin. There was no special reason to choose these particular derivatives. All nonvanishing derivatives give distinct solutions.
\begin{align}\label{d1}
\Psi_2=\frac{m^2\pi}{4}\{{\mathfrak K}_1,\,0,\,it_-{\mathfrak K}_2,\,-ix_+{\mathfrak K}_2\}.
\end{align}
\begin{align}\label{d3}
\Psi_4=\frac{m^3\pi}{24}\{-ix_-{\mathfrak K}_2,\,it_-{\mathfrak K}_2,\,2x_-t_-{\mathfrak K}_3,\,((a+it)^2-x^2-y^2+z^2){\mathfrak K}_3\}.
\end{align}
\begin{align}\label{d5}
\Psi_6=\frac{m^4\pi}{96}\{x_-t_-{\mathfrak K}_3,\,-((a+it)^2-x^2-y^2+z^2){\mathfrak K}_3,
4ix_-{\mathfrak K}_3/m-ix_-(x^2+y^2){\mathfrak K}_4,\,
4it_+{\mathfrak K}_3/m+t_-t_+^2{\mathfrak K}_4\}.
\end{align}
\begin{align}\label{d7}
\Psi_8=\frac{m^5\pi}{960}\{x_+t_-((a+it)^2-x^2-y^2+z^2){\mathfrak K}_5,-2x_+^2t_-^2{\mathfrak K}_5,\,
ix_+t_-^2{\mathfrak K}_4,\,ix_-^2t_-{\mathfrak K}_4\}.
\end{align}
\end{widetext}

\end{document}